# Beyond Batch Processing:
# Towards Real-Time and Streaming Big Data

*Saeed Shahrivari, and Saeed Jalili*

*Computer Engineering Department,*
*Tarbiat Modares University (TMU), Tehran, Iran*

*saeed.shahrivari@gmail.com, sjalili@modares.ac.ir*

**Abstract:** Today, big data is generated from many sources and there is a huge demand for storing, managing, processing, and querying on big data. The MapReduce model and its counterpart open source implementation Hadoop, has proven itself as the de facto solution to big data processing. Hadoop is inherently designed for batch and high throughput processing jobs. Although Hadoop is very suitable for batch jobs but there is an increasing demand for non-batch processes on big data like: interactive jobs, real-time queries, and big data streams. Since Hadoop is not proper for these non-batch workloads, new solutions are proposed to these new challenges. In this article, we discuss two categories of these solutions: real-time processing, and stream processing for big data. For each category, we discuss paradigms, strengths and differences to Hadoop. We also introduce some practical systems and frameworks for each category. Finally, some simple experiments are done to show effectiveness of some solutions compared to available Hadoop-based solutions.

## 1. Introduction

The "*Big Data*" paradigm is getting an expanding popularity recently. The "*Big Data*" term is generally used for datasets which are so huge that cannot be processed and managed using classical solutions like *Relational Data Base Systems (RDBMS)*. Besides volume, large velocity and variety are other challenges of big data. Numerous sources generate big data. Internet, Web, Online Social Networks, Digital Imaging, and new sciences like Bioinformatics, Particle Physics, and Cosmology are some example of sources for big data to be mentioned [1].

Until now, the most notable solution that is proposed for managing and processing big data is the MapReduce framework [2]. The MapReduce framework is initially introduced and used by Google. MapReduce offers three major features in a single package. These features are: simple and easy programming model, automatic and linear scalability, and built-in fault tolerance. Google announced its MapReduce framework as three major components: a MapReduce execution engine, a distributed file system called *Google File System (GFS)* [3], and a distributed NoSQL database called *BigTable* [4].

After Google's announcement of its MapReduce Framework, the Apache foundation started some counterpart open source implementation of the MapReduce framework. *Hadoop MapReduce* and *Hadoop YARN* as execution engines, the *Hadoop Distributed File System (HDFS),* and *HBase* as replacement for BigTable, were the three major projects [5]. Apache has also gathered some extra projects like: *Cassandara* a distributed data management system resembling to *Amazon Dynamo*, *Zookeeper* a high-performance coordination service for distributed applications, *Pig* and *Hive* for data warehousing, and *Mahout* for scalable machine learning.

From its inception, the Mapreduce framework has made complex large-scale data processing easy and efficient. Despite this, MapReduce is designed for batch processing of large volumes of data, and it is not suitable for recent demands like real-time and online processing. MapReduce is inherently designed for high throughput batch processing of big data that take several hours and even days, while recent demands are more centered on jobs and queries that should finish in seconds or at most, minutes.

In this article, we focus on two new aspects: real-time processing and stream processing solutions for big data. An example for real-time processing is fast and interactive queries on big data warehouses, in which user wants the result of his queries in less than seconds rather than in minutes or hours. Principally, the goal of real-time processing is to provide solutions that can process big data very fast and interactively. Stream processing deals with problems that their input data must be processed without being totally stored. There are numerous use cases for stream processing like: online machine learning, and continuous computation. These new trends need systems that are more elaborate and agile than the currently available MapReduce solutions like the Hadoop framework. Hence, new systems and frameworks have been proposed for these new demands and we discuss these new solutions.

We have divided the article into three main sections. First, we discuss the strength, features, and shortcomings of the standard MapReduce framework and its de facto open source implementation Hadoop. Then, we discuss real-time processing solutions. Afterwards, we discuss the stream processing systems. At the end of the article, we give some experimental results comparing the discussed paradigms. And finally, we give a conclusion.

## 2. The MapReduce framework

Essentially, MapReduce is a programming model that enables large-scale and distributed processing of big data on a set of commodity machines. MapReduce defines computation as two functions: *map* and *reduce.* The input is a set of key/value pairs, and the output is a list of key/value pairs. The *map* function takes an input pair and results a set of intermediate key/value pairs (which can be empty). The *reduce* function takes an intermediate key and a list of intermediate values associated to that key as its input, and results a set of final key/value pairs as the output. Execution of a MapReduce program involves two phases. In the first phase each input pair is given to a *map* function and a set of input pairs is produced. Afterwards in the second phase, all of the intermediate values that have the same key are aggregated into a list, and each intermediate key and its associated intermediate value list is given to a reduce function. More explanation and examples are available in [2].

The execution of a MapReduce program obeys the same two-phase procedure. Usually, distributed MapReduce is implemented using master/slave architecture. The master machine is responsible of assignment of tasks and controlling the slave machines. A schematic for execution of a MapReduce program is given in Figure 1. The input is stored over a shared storage like distributed file system, and is split into chunks. First, a copy of *map* and *reduce* functions' code is sent to all workers. Then, master assigns *map* and *reduce* tasks to workers. Each worker assigned a map task, reads the corresponding input split and passes all of its pairs to *map* function and writes the results of the *map* function into intermediate files. After the *map* phase is finished, the reducer workers read intermediate files and pass the intermediate pairs to *reduce* function and finally the pairs resulted by *reduce* tasks are written to final output files.

## 2.1. Apache Hadoop

There are several MapReduce-like implementations for distributed systems like *Apache Hadoop*, *Disco* from Nokia, *HPCC* from LexisNexis, *Dryad* from Microsoft [6], and *Sector/Sphere*. However, Hadoop is the most well-known and popular open source implementation of MapReduce. Hadoop uses master/slave architecture and obeys the same overall procedure like Figure 1, for executing programs. By default, Hadoop stores input and output files on its distributed file system, HDFS. However, Hadoop provides pluggable input and output sources. For example, it can also use NoSQL databases like HBase and Cassandra and even relational databases instead of HDFS.

Hadoop has numerous strengths. Some of its strengths come from the MapReduce model. For example, easy programming model, near-linear speedup and scalability, and fault tolerance are three major features. Besides these, Hadoop itself provides some extra features like: different schedulers, more sophisticated and complex job definitions using YARN, high available master machines, pluggable I/O facilities, and etc. Hadoop provides the basic platform for big data processing. For more usability, other solutions can be mounted over Hadoop [5]. Major examples are HBase for storing structured data on very large tables, Pig and Hive for data warehousing, and Mahout for machine learning.

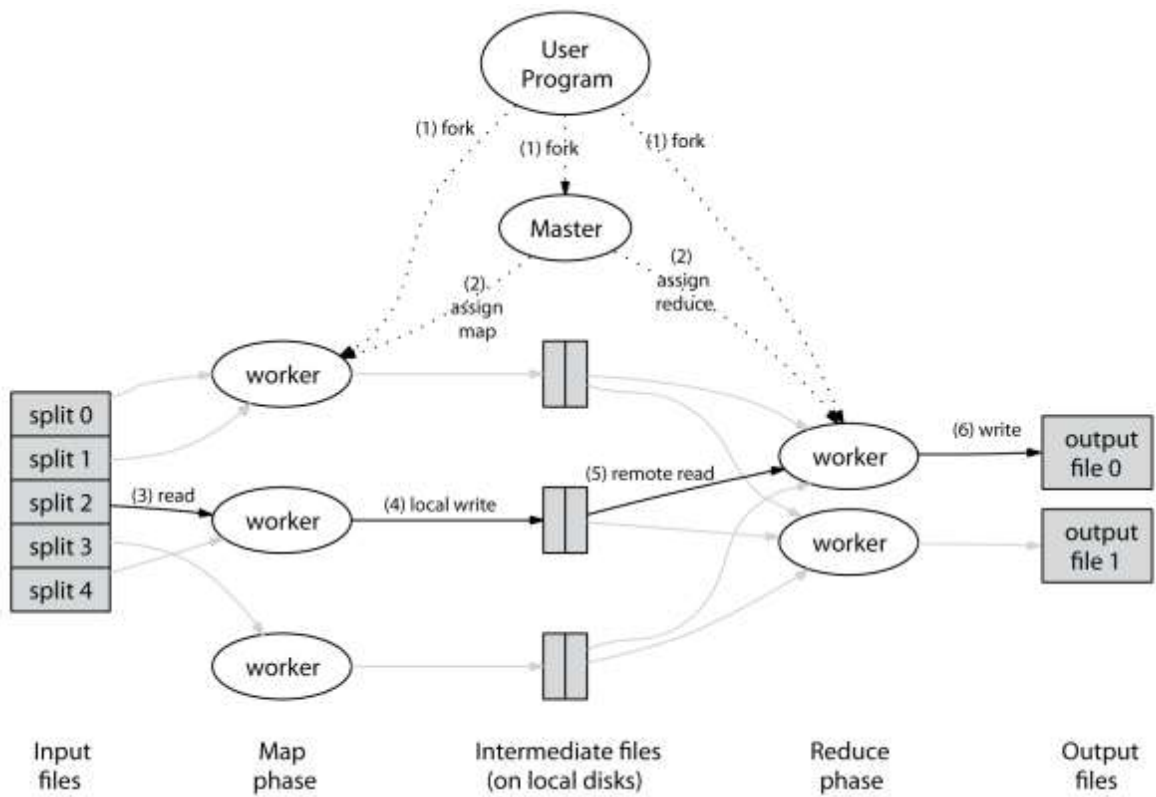

Figure 1: Execution of a MapReduce program [7]

Although Hadoop, i.e. standard MapReduce, has numerous strengths but it has several shortcomings too. MapReduce is not able to execute recursive or iterative jobs inherently [8]. Totally batch behavior is another problem. All of the input must be ready before job starts and this prevents MapReduce from online and stream processing use cases. The overhead of framework for starting a job, like copying codes and scheduling, is another problem that prevents it from executing interactive jobs and near real-time queries. MapReduce cannot run continuous computations and queries, too. These shortcomings have triggered creation of new solutions. Next, we will discuss two types of solutions: i) solutions that try to add real-time processing capabilities to MapReduce, and ii) solutions that try to provide stream processing of big data.

## 3. Real-time big data processing

Solutions in this sector can be classified into two major categories: i) Solutions that try to reduce overhead of MapReduce and make it faster to enable execution of jobs in less than seconds. ii) Solutions that focus on providing means for real-time queries over structured and unstructured big data using new optimized approaches. Here we discuss both categories respectively.

## 3.1. In-memory computing

Slowness of Hadoop is rooted in two major reasons. First, Hadoop was initially designed for batch processing. Hence, starting execution of jobs is not optimized for fast execution. Scheduling, task assignment, code transfer to slaves, and job startup procedures are not designed and programmed to finish in less than seconds. The second reason is the HDFS file system. HDFS by itself is designed for high throughput data I/O rather than high performance I/O. Data blocks in HDFS are very large and stored on hard disk drives which with current technology can deliver transfer rates between 100 and 200 megabytes per second.

The first problem can be solved by redesigning job startup and task execution modules. However, the file system problem is inherently caused by hardware. Even if each machine is equipped with several hard disk modules, the I/O rate would be several hundreds of megabytes per seconds. This means that if we store 1 terabytes of data on 20 machines, even a simple search over the data will take minutes rather than seconds. An elegant solution to this problem is *In-Memory Computing*. In a nutshell, in-memory computing is based on using a distributed main memory system to store and process big data in real-time.

Main memory delivers higher bandwidth, more than 10 gigabytes per second compared to hard disk's 200 megabytes per second. Access latency is also much better, nanoseconds versus milliseconds for hard disks. Price of RAM is also affordable. Currently, 1 TB of RAM can be bought with less than 20,000$. These performance superiority combined with dropping price of RAM makes in-memory computing a promising alternative to disk-based big data processing. There are few in-memory computing solutions available like: *Apache Spark* [9], *GridGain*, and *XAP*. Amongst them, Spark is both open source and free but others are commercial.

We must mention that in-memory computing does not means the whole data should be kept in memory. Even if a distributed pool of memory is available and the framework use that for caching of frequently used data, the whole job execution performance can be improved significantly. Efficient caching is especially effective when an iterative job is being executed. Both Spark and GridGain support this caching paradigm. Spark uses a primary abstraction called *Resilient Distributed Dataset (RDD)* that is a distributed collection of items [10]. Spark can be easily integrated with Hadoop and RDDs can be generated from data sources like HDFS and HBase. GridGain also has its own in-memory file system called GridGain File System (GGFS) that is able to work as either a standalone file system or in combination with HDFS, acting as a caching layer. In-memory caching can also help handling huge streaming data that can easily stifle disk-based storages.

Another important point to be mentioned is the difference between in-memory computing and in-memory databases and data grids. Although in-memory databases like Oracle *Times Ten* and VMware *GemFire* and in-memory data grids like *Hazelcast*, Oracle *Coherence,* and Jboss *Infinispan* are fast and have important use cases for today, but they differ from in-memory computing. In-memory computing is rather a paradigm than a product. As its name implies, in-memory computing deals with computing, too; in contrast to in-memory data solutions that just deal with data. Hence, it should also take care of problems like efficient scheduling, and moving code to data rather than wrongly moving data to code. Despite this, in-memory data grids can be used as a building block of in-memory computing solutions.

### 3.2. Real-time queries over big data

The first work in the area of solutions that try to enable real-time ad-hoc queries over big data is *Dremel* by Google [11]. Dremel use two major techniques to achieve real-time queries over big data: i) Dremel uses a novel columnar storage format for nested structures ii) Dremel uses scalable aggregation algorithms for computing query results in parallel. These two techniques enable Dremel to process complex queries in real-time. *Cloudera Impala* is an open source counterpart that tries to provide an open source implementation of Dremel techniques. For this purpose, Impala has developed an efficient columnar binary storage for Hadoop called *Parquet* and uses techniques of parallel DBMSs to process ad hoc queries in real-time. Impala claims considerable performance gains for queries with joins, over Apache Hive. Although Impala shows promising improvements over Hive, but still for long running analytics and queries, Hive is still a stable solution.

There are even more solutions in this sector. *Apache Drill* is also another Dremel-like solution. However, Drill is not designed to just be a Hadoop-only solution and it provides real-time queries against other storage systems like Cassandra. *Shark* is another solution that is built on top of Spark [12]. Shark is designed to be compatible with Apache Hive and it can execute all queries that are possible for Hive. Using in-memory computing capability and the fast execution engine of Spark, Shark claims up to 100x faster response times compared to Hive [12]. We should also mention the *Stinger* project by *Hortonworks* which is an effort to make 100x performance improvement and add SQL semantics to future versions of Apache Hive. The final mentionable solution is *Amazon Redshift* which is a propriety solution from Amazon that aims to provide a very fast solution to petabytes-scale warehousing.

### 4. Streaming big data

Data streams are now very common. Log streams, click streams, message streams, and event streams are some good examples. But, the standard MapReduce model and its implementations like Hadoop, is completely focused on batch processing. That is to say, before any computation is started, all of the input data must be completely available on the input store, e.g. HDFS. The framework process the input data and the output results are available only when all of the computation is done. On the other hand, a MapReduce job execution is not continuous. In contrast to these batch properties, today's applications need more stream-like demands in which the input data is not available completely at the beginning and arrives constantly. Also, sometimes an application should run continuously, e.g. a query that detects some special anomalies from incoming events.

Although MapReduce does not support stream processing, but MapReduce can partially handle streams using a technique known as *micro-batching*. The idea is to treat the stream as a sequence of small batch chunks of data. On small intervals, the incoming stream is packed to a chunk of data and is delivered to batch system to be processed. Some MapReduce implementations especially real-time ones like Spark and GridGain support this technique. However, this technique is not adequate for demands of a true stream system. Furthermore, the MapReduce model is not suitable for stream processing.

Currently, there are a few stream processing frameworks that are inherently designed for big data streams. Two notable ones are *Storm* from Twitter, and *S4* from Yahoo [13]. Both frameworks run on the Java Virtual Machine (JVM) and both process keyed streams. However, the programming model of the frameworks is different. In S4, a program is defined in terms of a graph of *Processing Elements (PE)* and S4 instantiates a PE per each key. On the other hand in Storm, a program is defined by two abstractions: *Spouts* and *Bolts*. A spout is a source of stream. Spouts can read data from an input queue or even generate data themselves. A bolt process one or more input streams and produces a number of output streams. Most of the process logic is expressed in bolts. Each Storm program is a graph of spouts and bolts which is called a *Topology* [14].

Considering the programming models we can say that in S4 the program is expressed for keys while in Storm the program is expressed for the whole stream. Hence, programming for S4 has a simpler logic while for Storm; programming is more complex but it is more versatile too. A major strength of Storm over S4 is fault tolerance. In S4, at any stage of process, if input buffer of a PE gets full, the incoming messages will be simply dropped. S4 also uses a check pointing strategy for fault tolerance. If a node crashes, its PEs will be restarted in another node form their latest state. Hence, any process after the latest state is lost [13]. However, Storm guarantees process of each tuple if the tuple successfully enters Storm. Storm does not store states for bolts and spouts but if a tuple does not traverse the Storm topology in a predefined period, the spout that had generated that tuple will replay it.

Actually, S4 and Storm take two different strategies. S4 proposes a simpler program model and it restricts the programmer in declaring the process but instead it provides simplicity and more automated distributed execution, for example automatic load balancing. In contrast, Storm gives the programmer more power and freedom for declaring the process but instead the programmer should take care of things like load balancing, tuning buffer sizes, and parallelism level for reaching optimum performance. Each of Storm and S4 has its own strengths and weaknesses, but currently Storm is more popular and has a larger community of users compared to S4. Due to demands and applications, Streaming big data will certainly grow much more in future.

## 5. Experimental results

We did some simple experiments for better illumination of the discussed concepts. The experiments are aimed to show the improvements of some of the mentioned systems compared to Hadoop. We do not intend to compare the performance of technologies in details. We just want to show that recent systems in the area of real-time big data are more suitable than Hadoop. We should mention that, stream processing has its own paradigms and it is not correct to compare stream processing solutions to Hadoop. Therefore, we did not consider stream processing solutions in experiments. For the case of real-time in-memory computing, we selected Spark. We executed simple programs like *WordCount* and *Grep*, and compared performance results to Hadoop. For the case of real-time queries over big data, a comprehensive benchmark is done by the Berkeley AMP Lab [15]. Hence, for this category we just reported a summary of that benchmark.

WordCount counts occurrences of each word in a given text and Grep extracts matching strings of a given pattern in a given text. WordCount is CPU intensive but Grep is I/O intensive if a simple pattern is being searched. We searched for simple word; hence, Grep is totally I/O intensive here. For this experiment, we used a cluster of 5 machines each having two 4-core 2.4 GHz Intel Xeon E5620 CPU and 20 GBs of RAM. We installed Hadoop 1.2.0, and Spark 0.8 on the cluster. For running WordCount and Grep, we used an input text file of size 40 GBs containing texts of about 4 million Persian web pages. For better comparison, we also executed the standard *wc* (version 8.13) and *grep* (version 2.10) programs of Linux on a single machine and reported their times, too. Actually, the wc command of Linux just counts number of all words, not occurrences of each word. Hence, it performs a much simpler job. The results are reported in Figure 2.

As diagrams show, Spark outperforms Hadoop in both experiments when input is on disk and when input is totally cached in RAM. In the WordCount problem, there is a little difference between in-memory Spark and on-disk Spark. On the other hand, for the Grep problem, there is a significant difference between in-memory Spark and on disk cases. Especially, when the input file is totally cached in memory, Spark executes Grep in a second while Hadoop takes about 160 seconds.

Berkeley AMP Lab has compared several frameworks and benchmarked their response time on different types of queries like: scans, aggregations, and joins on different data sizes. The benchmarked solutions are: Amazon Redshift, Hive, Shark, and Impala. The input data set is a set of HTML documents and two SQL tables. For better benchmarking, queries were executed with varying result sets: i) BI-like results which can be easily fit in a BI tool, ii) Intermediate results which may not fit in memory of a single node, and iii) ETL-like results which are so large that require several nodes to store.

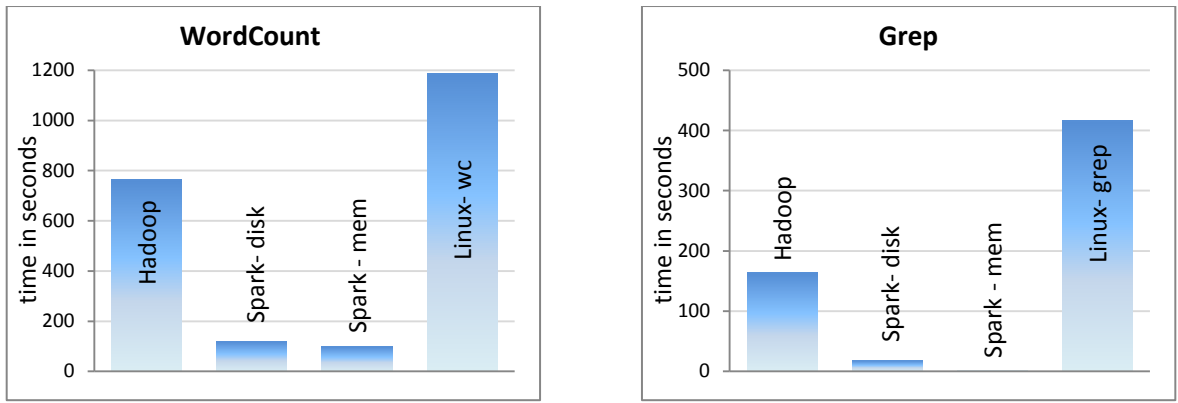

Figure 2: Hadoop performance compared to Spark

Two different clusters were used for this experiment: a cluster of 5 machines with total 342 GBs of RAM, 40 CPU cores, and 10 hard disks for running Impala, Hive, and Shark, and a cluster of 10 machines with total 150 GBs of RAM, 20 CPU cores, and 30 hard disks for executing Redshift. Both clusters were launched on the Amazon EC2 cloud computing infrastructure. The Berkeley AMP Lab benchmark is very comprehensive [15]. In this article, we just report the aggregation query results. The executed aggregation query is like "*SELECT * FROM foo GROUP BY bar*" SQL statement. The results are given in Figure 3.

As the results show, new-generation solutions show promising better performance compared to classic MapReduce-based solution, Hive. The only exception is the ETL-like query, in which Hive performs the same as Impala and this is because the result is so large that Impala cannot handle it properly. Despite this, other solutions like Shark and Redshift perform better than Hive for all result sizes.

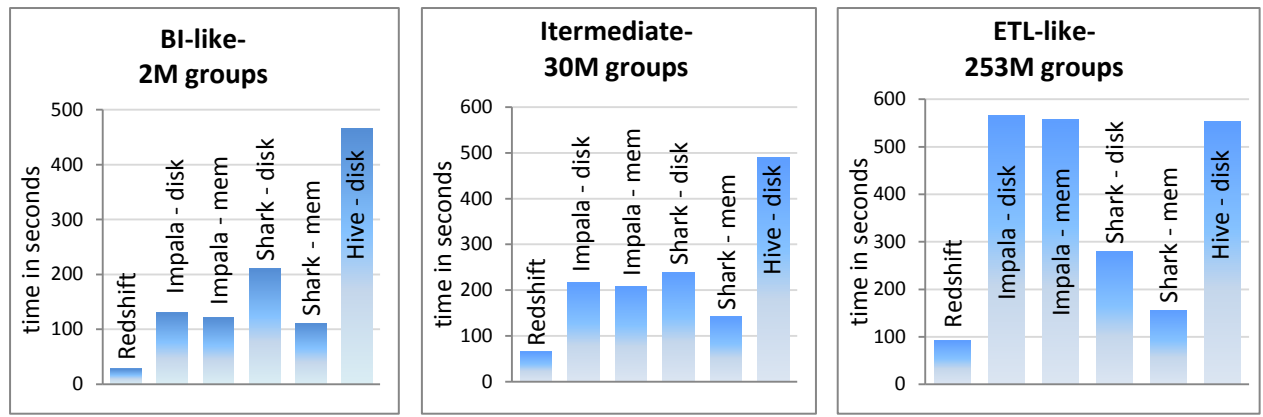

Figure 3: Berkeley AMP Lab's benchmark results for real-time queries

## 6. Conclusion

Big data has become a trend and some solutions have been provided for management and processing big data. The most popular solutions are MapReduce-based solutions and among them the Apache Hadoop framework is the most well-known. However, Hadoop is inherently designed for batch and high throughput job execution and it is suitable for jobs that process large volumes of data in a long time. In contrast, there are also new demands like interactive jobs, real-time queries, and stream data that cannot be handled efficiently by batch-based frameworks like Hadoop. These non-batch demands have resulted in creation of new solutions. In this article we discussed two categories: real-time processing, and streaming big data.

In the real-time processing sector, there are two major solutions: in-memory computing, and real-time queries over big data. In-memory computing uses a distributed memory storage that can be used either as a standalone input source or as a caching layer for disk-based storages. Especially, when the input totally fits in distributed memory or when the job has multiple iterations over input, in-memory computing can significantly reduce execution time. Solutions to real-time querying over big data mostly use custom storage formats and well-known techniques from parallel DBMSs for join and aggregation, and hence can response to queries in less than seconds. In the stream-processing sector, there are two popular frameworks: Storm, and S4. Each one has its own programming model, strengths and weaknesses. We discussed both frameworks and their superiority to MapReduce-based systems for stream processing.

We believe that, solutions to batch and high throughput processing of big data, like Hadoop, have reached to an acceptable maturity level. However, they are not suitable enough for non-batch requirements. Considering high demands for interactive queries and big data streams, in-memory computing shines as notable solution that can handle both real-time and stream requirements. Among discussed frameworks, Spark is a good example for this case which supports in-memory computing using RDDs, real-time and interactive querying using Shark, and stream processing using fast micro-batching. However, future will tell which approach will be popular in practice.